\documentstyle[prb,aps,epsf,rotate]{revtex}
\begin{document}
\draft

\twocolumn[\hsize\textwidth\columnwidth\hsize\csname@twocolumnfalse%
\endcsname

\title{Incommensurate phases in ferromagnetic spin-chains with weak
antiferromagnetic interchain interaction Chain}

\author{C. Pich}
\address{Department of Physics, University of California, Santa Cruz, 
CA 95064}

\author{F. Schwabl}
\address{Technische Universi\"at M\"unchen, James-Franck-Str., 
85747 M\"unchen, Germany\\
}
\date{\today}

\maketitle

\begin{abstract}
We study planar ferromagnetic spin-chain systems with weak
antiferromagnetic inter-chain interaction and dipole-dipole interaction. The
ground state depends sensitively on the relative strengths of antiferromagnetic
exchange and dipole energies $\kappa=J'a^2c/(g_L\mu_B)^2$. For increasing
values of $\kappa$, the ground state changes from a ferromagnetic via a
collinear antiferromagnetic and an incommensurate phase to a $120^o$ structure
for very large antiferromagnetic energy. Investigation of the magnetic phase 
diagram of the collinear phase, as realized in CsNiF$_3$, shows that the
structure of the spin order depends sensitivly on the direction of the magnetic
field in the hexagonal plane. For certain angular domains of the field  
incommensurate phases appear which are separated by commensurate phases. When
rotating the field, the wave vector characterizing the structure changes
continuously in the incommensurate phase, whereas in the commensurate phase the
wave vector is locked to a fixed value describing a two-sublattice
structure. This is a result of the competition between the exchange and the
dipole-dipole interaction.

\end{abstract}

\vskip 0.3 truein
]

\section{Introduction}
Ferromagnetic spin chains have been studied extensively, experimentally as well
as theoretically\cite{STEI72,STEI74,STEI76,STEI73,Shiba82,Shiba82a,Yamazaki80,Pich97}. 
Typical systems are CsNiF$_3$ and RbFeCl$_3$ which show quasi one-dimensional
behavior due to the small lattice constant along the $c$-axis. Nearest neighbor
spins couple with a ferromagnetic exchange interaction along the spin
chain. A planar anisotropy of the same order of magnitude is found because of
the non-vanishing orbital angular momentum of the magnetic ions. Perpendicular
to the chain axis the magnetic ions are located on a triangular (hexagonal)
lattice with a much larger lattice constant $a$, i.e. the magnetic ions form a
simple hexagonal lattice. Thus, at high temperatures a
pronounced one-dimensional behavior can be found experimentally. Planar
ferromagnets still have rotational symmetry, so that no long range order
exists. Nevertheless, spin waves were measured for ${\bf q}\ne
0$\cite{STEI73}. At low temperatures the systems undergo a phase transition to
a three-dimensional ordered phase for which the antiferromagnetic exchange
interaction and the dipole-dipole interaction are responsible. The precise
structure of the order in the hexagonal plane depends sensitively on the
relative strengths of these two competing interactions. RbFeCl$_3$ has a
nearly $120^o$ structure because the exchange energy is large. In contrast,
CsNiF$_3$ has a collinear orientation of the spins due to the large dipole
energy. Up to now the three-dimensional ordering has not attracted much 
interest, but interesting behavior is expected due to the competing
interactions which can lead to frustration.

Recently Baehr et al.\cite{baehr} have measured the magnetic excitations in
the three-dimensional ordered CsNiF$_3$ ($T<T_N=2.7$ K) by inelastic neutron
scattering. For the first time the magnitude of the antiferromagnetic exchange
energy in the plane has been determined by measuring the dispersion
relation. The important role of the dipole-dipole interaction is manifested by
the discontinuity in the magnon energy at the zone center ({\bf q}=0). This is
an effect of the long range interaction. In this paper we want to study the
whole range from vanishing exchange energy up to the region where the dipole
energy is small. It will be shown that instead of the frustrated $120^o$
structure an incommensurate phase will be established. The dispersion relations
are evaluated for the commensurate phases. In a second step we investigate the
magnetic phase diagram for the collinear phase, especially for CsNiF$_3$, and
calculate the critical fields. The Ni atoms form a simple hexagonal lattice
with lattice constants $a=b=6.21$\AA\, and $c=5.2$\AA\, the Land\'e factor
$g_L=2.25$\cite{STEI91} and the spin $S=1$. The outline of the paper is as
follows. In chapter II we introduce the model, in chapter III the ground state
for vanishing field strength is determined and in chapter IV we study the
excitations of the ferromagnetic and the collinear phase. In chapter V the
collinear antiferromagnetic phase with finite fields in the hexagonal plane is 
investigated. In the appendix we summarize the main relations for the dipole
energy.

\section{Model}
The starting point of our investigation is the Heisenberg Hamiltonian 
\begin{eqnarray}
  H &=& -2J\sum_{i}{\bf S}_i{\bf S}_{i+1}+A\sum_{l} {(S_l^z)}^2 - \nonumber\\
&&
  \sum_{\alpha,\beta}\sum_{l,l'}\left(J_{ll'}'\delta^{\alpha\beta}+ 
    A^{\alpha\beta}_{ll'}\right)S_l^\alpha S_{l'}^\beta -g_L\mu_B {\bf
    H}_0\sum_l {\bf S}_l \ . 
  \label{hamilton}
\end{eqnarray}
Here $J$ denotes the ferromagnetic nearest-neighbor intrachain interaction, $A$
the single-ion anisotropy, $J_{ll'}'$ the interchain and
$A^{\alpha\beta}_{ll'}$ the dipole-dipole interaction
\begin{equation}
  A^{\alpha\beta}_{l,l'} = -{(g_L\mu_B)^2\over
    2}\left\{{\delta^{\alpha\beta}\over 
      |{\bf x}_l -{\bf x}_{l'}|^3}+{3({\bf x}_l -{\bf x}_{l'})^\alpha ({\bf
      x}_l -{\bf x}_{l'})^\beta \over |{\bf x}_l -{\bf x}_{l'}|^5}\right\}.
\label{Dipoltensor}
\end{equation}
$i$ indicates positions on one and the same spin chain, whereas $l$ indicates
all spin positions. The notation for $x_l$, used here, is given in the
appendix. ${\bf H}_0$ is an external field perpendicular to the chain
axis. We are interested in the case of a planar ferromagnetic chain,
i.e. $J,A>0$, which means that the spins are forced to lie in the hexagonal
plane perpendicular to the chain.  The following calculations are valid only
for weak inter-chain interactions, i.e. $J,A\gg J',(g_L\mu_B)^2/a^2c$, where
$a$ indicates the lattice constant of the triangular lattice in the plane and
$c$ the lattice constant of the chain. Because of the long-range of the
dipolar interaction, summations over the whole system are performed best 
by means of the Ewald summation technique \cite{Bonsal77} (s. appendix). In
general, we can differentiate between one-dimensional behaviour at high
temperatures due to the large intra-chain energy and the three-dimensional
properties at low temperatures.

\section{Ground states for ${\bf H}_0=0$}

\begin{figure}
\centerline{
 \epsfxsize=0.8\columnwidth\rotate[r]{\epsfbox{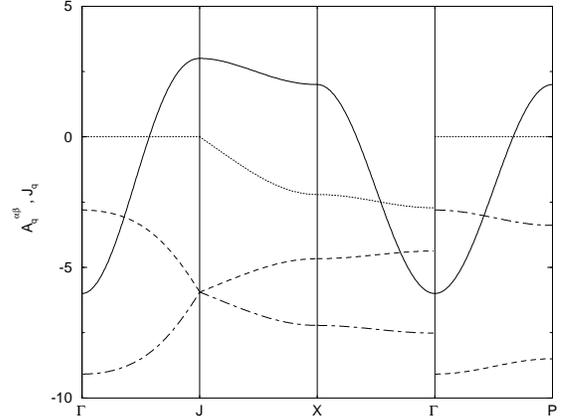}}
}
\caption{Antiferromagnetic exchange energy $J_{\bf q}'$ and
  dipolar interaction $A^{\alpha\beta}_{\bf q}$ for selected directions in the
  Brillouin zone. The solid line represents the isotropic exchange, the dashed
  line $A^{yy}_{\bf q}$, the dot-dashed line $A^{xx}_{\bf q}$ and the dotted
  line $A^{xy}_{\bf q}$. The dipole components are given in units
  $(g_L\mu_B)^2/a_3$ and the exchange in units $J'$.}
\label{Dipol}
\end{figure}

First we study the ground state of the three-dimensional system
without an external field. Due to the pronounced one-dimensional behaviour the
classical ground state consists of ferromagnetic spin-chains with the spins
ordered in the plane perpendicular to the chains. The spin structure within the
hexagonal plane then depends only on the antiferromagnetic exchange and the
dipolar interaction. For the following investigation we introduce a
dimensionless parameter $\kappa$, which is the ratio of the antiferromagnetic
exchange to the dipolar interaction ($\kappa_{\rm CsNiF_3}=0.79$):
\begin{equation}
\kappa = {J'a^2c\over (g_L\mu_B)^2}\, .
\end{equation}
Fourier transformation of the Hamiltonian (Eq. (\ref{hamilton})) yields  
\begin{equation}
  H = -\sum_{\alpha,\beta}\sum_{\bf q} \left( J_{\bf q}\delta^{\alpha\beta}-
    A\delta^{\alpha z}\delta^{\beta z}+J_{\bf q}'\delta^{\alpha\beta}+
    A^{\alpha\beta}_{\bf q}\right)S_{\bf q}^\alpha S_{-{\bf q}}^\beta \, ,
\end{equation}
with the nearest-neighbor exchange energies (intrachain and interchain,
$J,J'>0$) 
\begin{eqnarray}
  J_{\bf q} & = & 2J\cos q_z \\ J_{\bf q}'& = & -2J'\left(\cos q_x
    +2\cos{q_x\over 2}\cos{\sqrt 3q_y\over 2}\right) \, .
\end{eqnarray}
The classical ground state is determined by the maximal eigenvalue of 
\begin{equation}
V^{\alpha\beta}_{\bf q} = \left( J_{\bf q}'\delta^{\alpha\beta}+
A^{\alpha\beta}_{\bf q}\right)\, , 
\label{V_0}
\end{equation}
with the wave-vector ${\bf q}$ constrained to the plane ($q_z=0$). The
eigenvalues of the three by three matrix (Eq. \ref{V_0}) are given by
\begin{eqnarray}
\lambda_1 & = & V^{zz}_{\bf q} \nonumber\\
\lambda_{2/3} & = & {1\over 2}\left( V^{xx}_{\bf q}+V^{yy}_{\bf q}\pm
\sqrt{(V^{xx}_{\bf q}-V^{yy}_{\bf q})^2+4\left(V^{xy}_{\bf q}\right)^2 }\,
\right)\, , 
\end{eqnarray}
because the off-diagonal components $A^{xz}_{\bf q}$ and $A^{yz}_{\bf q}$
vanish for wave vectors in the plane. An out-of-plane orientation of the spins
increases the energy ($V^{zz}_{\bf q}$) because of the planar anisotropy. Thus,
the wave vector, which maximizes $\lambda_2$, describes the classical ground
state. In Fig. (\ref{Dipol}) the exchange energy $J_{\bf q}$ and the the three
components of the symmetrical dipole tensor $A^{\alpha\beta}_{\bf q}$ are
plotted ($a/c=2.39$ and $J'=1$). The maximal value for pure antiferromagnetic
exchange is found to be for ${\bf q}_0= 4\pi/3a(1,0,0)$, i.e. a 120$^o$
structure.
For pure dipolar interaction the minimal value is at wave vector ${\bf q} =0$,
a  ferromagnetic orientation in correspondance to the two-dimensional
counterpart, the triangular lattice \cite{Pich94}. The ferromagnetic structure
is stable as long as the antiferromagnetic exchange energy is lower than the
dipolar interaction leading to   
\begin{equation}
-8J'< A^{11}_{0}-A^{11}_{{\bf q}_1}\qquad\qquad {\bf q}_1=2\pi/\sqrt 3a(0,1,0)
\, .
\label{stab1}
\end{equation}

Due to the semiconvergence of the dipole sums for a ferromagnetic structure
(${\bf q}= 0$), the dipole components are not analytic at the zone center. For
spherically shaped systems the values for the dipole tensor are given in the
appendix. For CsNiF$_3$ Eq. (\ref{stab1}) reduces to the inequality
\begin{equation}
J'< 7\  mK\, \qquad \kappa < \kappa_{c1} = 0.20
\label{kapp-fm}
\end{equation}
As a result of the hexagonal symmetry the ferromagnetic ground state is
continuously degenerate. 

\begin{figure}
\centerline{
 \epsfxsize=0.6\columnwidth\rotate[r]{\epsfbox{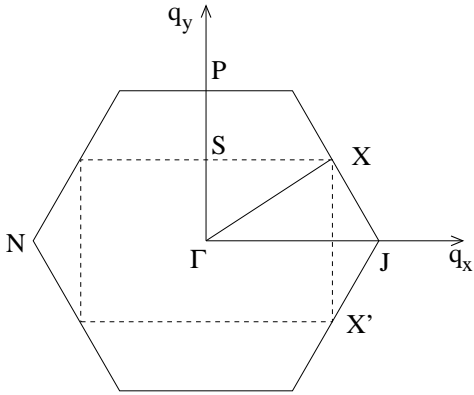}}
}
\vskip .5cm
\caption{The Brillouin zones of the hexagonal plane. The
  hexagon is the crystallographic and the rectangle (dashed) is the
  magnetic one. Included are the three directions for the scattering
  measurement which are realized in the domains. }
\label{BrillFig}
\end{figure}

When the exchange energy exceeds the above inequality (Eq. (\ref{stab1})) the
classical ground state changes to a collinear antiferromagnetic spin
orientation \cite{baehr}. There the spins are orientated along 
the lattice axes so that there exist only three discrete states called
domains (Fig. \ref{Spinstruct0}). Domain A can be described by the wave vector
${\bf q}_1$, domains C 
and B by  ${\bf q}_2=\pi(1,1/\sqrt 3,0)$ and ${\bf q}_3=\pi(1,-1/\sqrt 3,0)$ 
respectively. These wave vectors correspond to points $P$, $X$ and $X'$ in the
Brillouin zone (Fig. (\ref{BrillFig})).

\begin{figure}
\centerline{
 \epsfxsize=0.6\columnwidth{\epsfbox{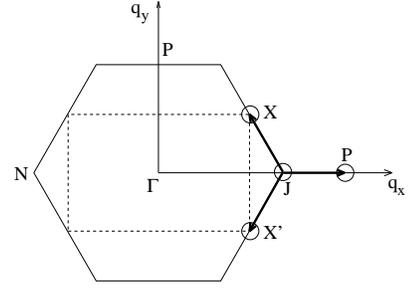}}
}
\vskip .5cm
\caption{Transition from the collinear antiferromagnetic
  phase, points $P$, $X$ and $X'$, towards the three-sublattice phase $J$. The
  intermediate phases are incommensurate, i.e. the wave vectors, defining the
  structure, have no commensurate relation to the periodicity of the underlying
  lattice.}
\label{inc-domains}
\end{figure}

For increasing exchange energy the collinear ground state does not change to a
frustrated $120^o$ structure as one might have assumed. Rather, the system
changes to an incommensurate phase. This results from the fact that the dipole
energy depends linearly on the wave vector near point $J$, in contrast to a
quadratic behavior for the exchange energy (Fig. \ref{Dipol}). Therefore, the
wave vector for the lowest ground state energy changes continuously from point
$J$ towards point $J$, which would correspond to a three sublattice state.
Point $J$ is reached only for vanishing dipole energy. Due to the hexagonal
symmetry the same situation holds for points $X$ and $X'$, i.e. the
three-domain structure from the collinear phase survives in the incommensurate 
region\cite{Shiba82,Shiba82a}. The three domains and the route to the three 
sublattice structure can be seen in Fig. (\ref{inc-domains}).

\begin{figure}
\centerline{
 \epsfxsize=0.8\columnwidth\rotate[r]{\epsfbox{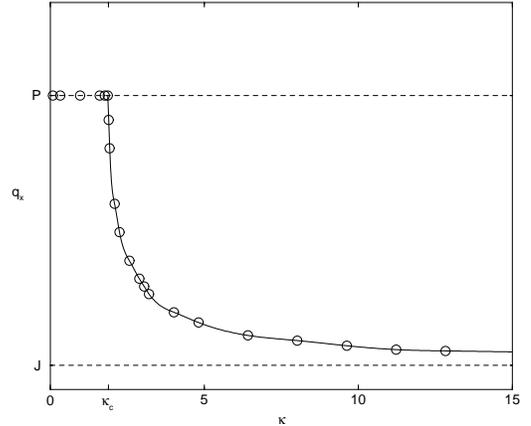}}
}
\vskip .5cm
\caption{Incommensurate phase. For $\kappa < \kappa_c$
  the collinear phase ($P$) is established, and for larger values the
  continuous change of the wave vector, describing the intermediate phase, is
  plotted. The three-sublattice phase ($J$) is reached only asymptotically. The
  same route holds for the other two domains.}
\label{incommensurate}
\end{figure}

The incommensurate wave vector behaves near $P$ as 
\begin{equation}
{\bf q}_c \simeq {\bf q}_1 -\sqrt{{\kappa - 1.88}\over 7\kappa/24
-0.32}(1,0,0)\, ,
\end{equation}
i.e. a square root dependence. For increasing exchange energy point $J$ is
reached asymptotically (s. Fig. (\ref{incommensurate})). 
Near point $J$ the
wave vector can be approximated by
\begin{equation}
{\bf q}_c\simeq {\bf q}_0 +1.32\kappa(1,0,0)\, .
\label{inc-wv}
\end{equation}

For systems with a dipolar strength as in CsNiF$_3$, the transition to the
incommensurate phase yields 
\begin{equation}
J' =  59\  mK\, \qquad \kappa = \kappa_c =1.88\, .
\label{coll-bound}
\end{equation}

\section{Excitations}
Now we investigate the spin wave frequencies for the two commensurate
phases  within linear spin wave theory. The incommensurate phase resists such
an analysis because of the infinite primitive cell. 

In linear spin wave theory the frequencies of the magnons can be derived via
the the Holstein-Primakoff transformation (HP) which transforms the spin
operators to Bose operators. Therefore, Bose operators $a_l$ and
$a_l^{\dag}$\cite{Ziman69} are introduced, which are given up to bilinear order
by  
\begin{mathletters}
\begin{eqnarray}
 \tilde S_l^z  &=&  (S-a_l^\dagger a_l), \label{trafo1}\\
 \tilde  S_l^x &=&  \sqrt{S\over 2}(a_l+a_l^\dagger),\label{trafo2}\\
 \tilde  S_l^y  &=&  - i\sqrt{S\over 2}(a_l-a_l^\dagger)\label{trafo3}\, 
\end{eqnarray}
\end{mathletters}
with the local spin vector $\tilde{\bf S}_l$. However, first the classical
ground state must be determined, and then the the above expressions have to be
inserted so that $\tilde S_l^z$ corresponds to the local $z$-component of each
spin.

\subsection{Ferromagnetic phase}
In this phase all spins are aligned ferromagnetically in the hexagonal plane. 
We choose the $z$-component of the spins along the $x$-axis. After HP
transformation the Hamilton operator has the form
\begin{equation}
  H  = E_{fm}^{cl}+ \sum_{\bf q}A_{\bf q}~a_{\bf
    q}^{\dagger}a_{\bf q} + {1\over 2} 
  B_{\bf q}~(a_{\bf q}~a_{-\bf q}+a_{\bf q}^{\dagger}a_{-\bf
    q}^{\dagger})
\label{H_ferro}
\end{equation}
with the coefficients
\begin{eqnarray}
  A_{\bf q} & = & SA+2S(J_{0}-J_{\bf q})+2S(J_{0}'-J_{\bf q}')+\nonumber\\
&&   S(2A^{xx}_0-A^{yy}_{\bf q}-A^{zz}_{\bf q})\\ 
  B_{\bf q} & = & -SA+ S(A^{zz}_{\bf q}-A^{yy}_{\bf q}+2iA^{yz}_{\bf q})\, 
\end{eqnarray}
and the classical ground state energy
\begin{equation}
E_{fm}^{cl} = -NS^2 (J_0+J'_0+A_0^{xx})\, .
\label{fm}
\end{equation}
The planar anisotropy $A$ does not contribute to this energy. The spin wave
frequency following Eq. (\ref{H_ferro}), given by 
\begin{equation}
E_{\bf q} = \sqrt{A_{\bf q}^2-|B_{\bf q}|^2}\, .
\end{equation}
Due to the rotation symmetry in the plane
the magnon frequency  vanishes at the zone center. This spectrum becomes
unstable for increasing exchange energy $J'$ at point P which corresponds to a
transition to the collinear antiferromagnetic phase (s. section III).

\subsection{Collinear antiferromagnetic phase}
\begin{figure}
\centerline{
 \epsfxsize=0.6\columnwidth\rotate[r]{\epsfbox{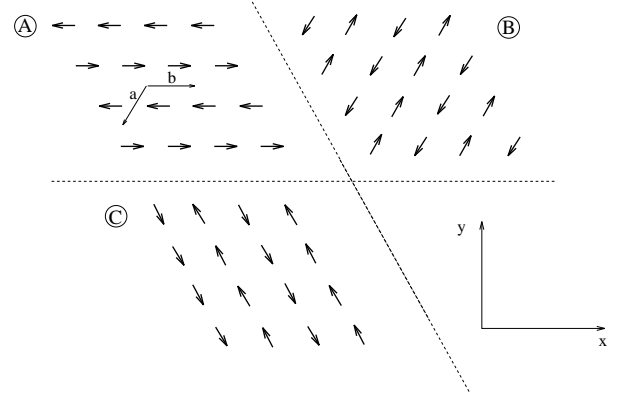}}
}
\vskip .5cm
\caption{The ground state for CsNiF$_3$ in the
  hexagonal plane is one of the three (A--C) shown configuration
  called domains. In domain A the two primitive vectors are
  represented. The antiferromagnetic modulation can be described by ${\bf
  q}_1$, ${\bf q}_2$ and ${\bf q}_3$ for domain A, B and C respectively. }
\label{Spinstruct0}
\end{figure}

\begin{figure}
\centerline{
 \epsfxsize=0.8\columnwidth{\epsfbox{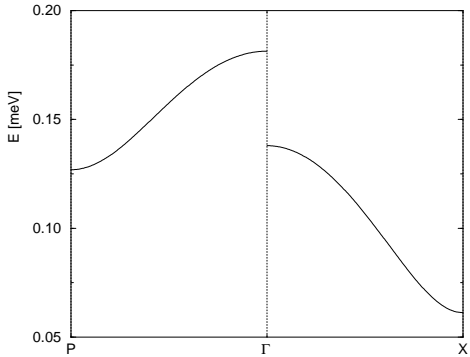}}
}
\vskip .5cm
\caption{Dispersion relation for CsNiF$_3$ (vanishing field)
  for two directions. There is a discontinuity at the zone center.}
\label{dispersion}
\end{figure}

In the ground state there exist three domains (Fig. (\ref{Spinstruct0})). Spin
wave theory is applied to domain A, where the spins are orientated along the
$x$-axis. The antiferromagnetic modulation is given by the wave vector ${\bf
q}_1={2\pi\over a\sqrt 3}(0,1,0)$. For domains C and B the analogous wave
vectors are ${\bf q}_2={\pi\over a}(1,1/\sqrt 3,0)$ and ${\bf
q}_3={\pi\over a}(1,-1/\sqrt 3,0)$ respectively. 
The Holstein-Primakoff transformation can be applied for the 
crystallographic Brillouin zone\cite{Pich93} by using the factor $e^{i{\bf
q}_1{\bf x}_l}$ for the antiferromagnetic modulation 
\begin{eqnarray}
 H  &= & E_g^{cl}+\sum_{\bf q}\left\{A_{\bf q}~a_{\bf
q}^{\dagger}a_{\bf q} + {1\over 2} 
B_{\bf q}~(a_{\bf q}~a_{-\bf q}+a_{\bf q}^{\dagger}a_{-\bf
q}^{\dagger})+\right. \nonumber \\
&& \left.~~ C_{\bf q}~a_{\bf q}~a_{-{\bf q}-{\bf q}_1}+
C_{\bf q}^{\ast}a_{\bf q}^{\dagger}a_{-{\bf q}-{\bf q}_1}^{\dagger}+\right.
\nonumber\\ 
&& \left.D_{\bf q}~a_{\bf q}^{\dagger}a_{{\bf q}+{\bf q}_1}+D_{\bf q}^{\ast}~
a_{{\bf q}+{\bf q}_1}^{\dagger}a_{\bf q}\right\}\label{H_afm}
\label{H_coll}
\end{eqnarray}
with the coefficients
\begin{eqnarray}
A_{\bf q}  & = & S(2J_{{\bf q}_1}-J_{\bf q}-J_{{\bf q}+{\bf q}_1})+SA+
\nonumber\\ 
&& S(2J_{{\bf q}_1}'-J_{\bf q}'-J_{{\bf q}+{\bf q}_1}')+
S(2A^{xx}_{{\bf q}_1}-A^{yy}_{\bf q}-A^{zz}_{{\bf q}+{\bf q}_1})\\ 
B_{\bf q}  & = & S(J_{{\bf q}+{\bf q}_1}-J_{\bf q})-SA+\nonumber\\
&& S(J_{{\bf q}+{\bf q}_1}'-J_{\bf q}')+
S(A^{zz}_{{\bf q}+{\bf q}_1}-A^{yy}_{\bf q})\nonumber \\
C_{\bf q}  & = &  iSA^{yz}_{\bf q}\\
D_{\bf q} & = & iSA^{yz}_{\bf q}
\end{eqnarray}
and the ground state energy
\begin{equation}
E_g^{cl} = -NS^2(J_0+J_{{\bf q}_1}' +A^{xx}_{{\bf q}_1})\, . 
\label{Eg}
\end{equation}
Diagonalization of the quadratic Hamiltonian (Eq. (\ref{H_coll})) can be done
by a generalized Bogoljubov transformation\cite{Pich93}. The magnon frequency
is given by 
\begin{equation}
E_{\bf q}^{(1/2)} = \sqrt{{1\over 2}(\Omega_1 \pm \Omega_2)}
\end{equation}
with
\[
\Omega_1 =A_{\bf q}^2-B_{\bf q}^2+A_{{\bf q}+{\bf q}_1}^2-B_{{\bf q}+{\bf 
q}_1}^2 +8C_{\bf q}~C_{{\bf q}+{\bf q}_1}
\]
and
\[\Omega_2^2 = (A_{\bf q}^2-B_{\bf q}^2-A_{{\bf q}+{\bf q}_1}^2+B_{{\bf q}+
{\bf q}_1}^2)^2+
\]
\[
16[C_{{\bf q}+{\bf q}_1}(A_{{\bf q}+{\bf q}_1}-B_{{\bf q}+{\bf q}_1})-C_{\bf
q}~ (A_{\bf q}-B_{\bf q})]
\]
\[ \times [C_{\bf q}~(A_{{\bf q}+{\bf q}_1}+B_{{\bf q}+{\bf q}_1})-
C_{{\bf q}+{\bf q}_1}(A_{\bf q}+B_{\bf q})]\, .\]
Here we changed to the magnetic Brillouin zone, which is half the
crystallographic, and therefore two branches appear. Note that the dipole
energy lifts the degeneracy for $q_z\ne 0$ because of the off-diagonal dipole
component $A^{yz}_{\bf q}$. The splitting of the two branches is of the order
of the dipole energy. For wave vectors in the plane the off-diagonal component
vanishes and the two magnon branches are degenerate. For wave vectors in the
hexagonal plane ($q_z=0$) stability of the collinear phase requires
\[
A_{\bf q} > 0, \qquad A_{\bf q} > |B_{\bf q}|, \qquad {\bf q} = (q_x,q_y,0)\,
.\]
From these inequalities we recover the upper bound for the exchange energy $J'$
(Eq. (\ref{coll-bound})). Note that for $J$ and $A$ much larger than $J'$ and
the dipole energy the inequalities are independent of $J$ and $A$. In
Fig. (\ref{dispersion}) the dispersion relation is plotted\cite{baehr} for
CsNiF$_3$ in two different directions of the Brillouin zone. 
Note the discontinuity at the zone center, which is a result of the nonanalytic
behavior of the dipole-dipole interaction (s. appendix).

\section{Collinear antiferromagnetic phase with ${\bf H}_0\ne 0$}

In this section we study the ground state and the dispersion relation
for the collinear antiferromagnetic phase in a homogeneous external magnetic
field. First we investigate the N\'eel phase and find that it is stable in
the presence of a longitudinal field (parallel to the spin orientation) up to a
critical value $H_0^c$. A transverse field always leads to a reorientation of
the spins, i.e. a canted structure. In a second section we study the
paramagnetic phase; especially we derive the instability boundary for arbitrary
field direction at which the system undergoes a transition to a canted
structure. Thus we obtain the corresponding wave vectors characterizing the
intermediate phases. In the last section we derive the stability equations for
a general two-sublattice structure and the conditions under which conventional
spin-flop phases can occur.

\subsection{N\'eel phase}

We consider domain A with the magnetic field along the $x$-axis. The classical
ground state energy (Eq. (\ref{Eg})) is not affected by the magnetic field
because of the alternation of the spins. In the bilinear part of the
Hamiltonian (Eq. (\ref{H_coll})) only the coefficient $D_{\bf q}$ changes to 
\begin{equation}
D_{\bf q} =  iSA^{yz}_{\bf q}+{1\over 2}g_L\mu_B H_0\, .
\end{equation}
This affects the two branches of the dispersion relation in the following way:
\begin{equation}
E_{\bf q}^{(1/2)} = \sqrt{{1\over 2}(\Omega_1 \pm \Omega_2)}
\end{equation}
with
\[
\Omega_1 =A_{\bf q}^2-B_{\bf q}^2+A_{{\bf q}+{\bf q}_1}^2-B_{{\bf q}+{\bf 
q}_1}^2 +8C_{\bf q}~C_{{\bf q}+{\bf q}_1}+2(g_L\mu_B H_0)^2
\]
and
\[\Omega_2^2 = (A_{\bf q}^2-B_{\bf q}^2-A_{{\bf q}+{\bf q}_1}^2+B_{{\bf q}+
{\bf q}_1}^2)^2+
\]
\[
16[C_{{\bf q}+{\bf q}_1}(A_{{\bf q}+{\bf q}_1}-B_{{\bf q}+{\bf q}_1})-C_{\bf
q}~ (A_{\bf q}-B_{\bf q})]
\]
\[ \times [C_{\bf q}~(A_{{\bf q}+{\bf q}_1}+B_{{\bf q}+{\bf q}_1})-
C_{{\bf q}+{\bf q}_1}(A_{\bf q}+B_{\bf q})]\]\[+4(g_L\mu_B H_0)^2((A_{\bf
q}+A_{{\bf q}+{\bf q}_1})^2-(B_{\bf q}-B_{{\bf q}+ {\bf q}_1})^2)\, .
\]
The effect of the magnetic field is to lower (lift) the energy of the lower
(upper) branch. The collinear ground state is stable as long as the second mode
$E_{\bf q}^{(2)}$ is positive. The energy of this mode vanishes with increasing
field strength at first at the boundary of the magnetic Brillouin zone at
${\bf q}_4 = {\pi\over a}(1,0,0)$. This implies a critical field of
\begin{equation}
  g_L\mu_BH^c_0 = 2S\sqrt{(A_{{\bf q}_1}^{xx}-A_{{\bf q}_4+{\bf
  q}_1}^{yy}) (A_{{\bf q}_1}^{xx}-A_{{\bf q}_4}^{yy})}\, .
\label{H-crit}
\end{equation}
Surprisingly, the critical field depends only on the dipolar energy because 
the antiferromagnetic exchange energy obeys $J_{{\bf q}_4}'=J_{{\bf q}_4+{\bf
q}_1}'= J_{{\bf q}_1}'$. The wave vector ${\bf q}_4$ describes a
four-sublattice system, i.e. the primitive cell of the magnetic lattice
consists of four spins. For CsNiF$_3$ Eq. (\ref{H-crit}) yields for the
critical field
\begin{equation}
  H^c_0 = 47 {\rm mT}\ .
\end{equation}
Above this value the collinear structure becomes unstable and an intermediate
state occurs, which is investigated in the next section. The critical value is
proportional to the energy gap at wave vector ${\bf q}_4$ present without field
which then is increasingly lowered by applying a field. This phenomenon is
already known for systems with an easy-axis anisotropy and a magnetic field
parallel to the spins\cite{Keffer66}.

\subsection{Paramagnetic Phase}

\begin{figure}
\centerline{
 \epsfxsize=0.9\columnwidth{\epsfbox{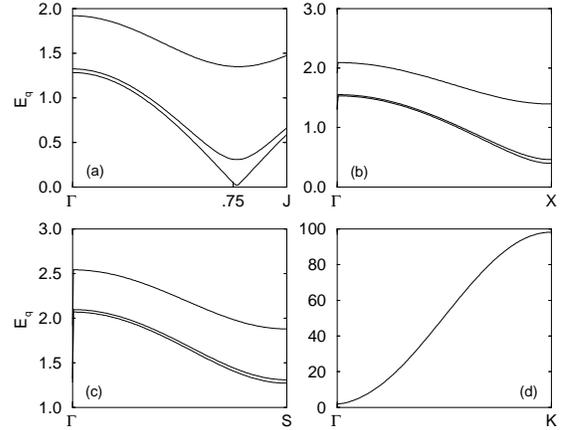}}
}
\vskip .5cm
\caption{Dispersion relation for CsNiF$_3$ in the paramagnetic
  phase. The field is aligned along the $x$-axis ($\varphi=0$). The upper
  curves are calculated for $H_0=2.1H_0^{cx}$, the middle curves for $H_0=1.06
  H_0^{cx}$ and the lower curves are just above the critical field
  $H_0^{cx}$. Point K equals ${\bf q}={\pi\over c}(0,0,1)$. In graph (a) one
  can see the softening of the dispersion at an incommensurate wave vector.}
\label{disp-para}
\end{figure}

Before studying the noncollinear phase, i.e. the spin-flop-like phase, it is
instructive to investigate the paramagnetic phase. This phase is established
for strong magnetic fields in the hexagonal plane so that all spins align
parallel to it. When the magnetic field strength is lowered, the paramagnetic
phase becomes unstable and a transition to a canted structure will occcur. From
this instability point we obtain the wave vector characterizing the canted
structure. Because the ground state of CsNiF$_3$ is not invariant with respect
to a rotation around the spin-chain axis (recall that there are three domains
A--C), the direction of the field plays a crucial role. Thus, we get a
non-circular instability line, i.e. the absolute value of the critical field at
which the paramagnetic phase becomes unstable depends on the angle the magnetic
field encloses with the $x$-axis. In consequence to the hexagonal symmetry it
is sufficient to study a range of 60$^o$. In the following $\varphi$ denotes
the angle between the external magnetic field ${\bf H}_0$ and the $x$-axis.
After a HP transformation the Hamiltonian reads: 
\begin{equation}
  H  = E_{PM}^{cl}+ \sum_{\bf q}A_{\bf q}~a_{\bf
    q}^{\dagger}a_{\bf q} + {1\over 2}\left( 
    B_{\bf q}a_{\bf q}~a_{-\bf q}+B_{\bf q}^\ast a_{\bf q}^{\dagger}a_{-\bf
    q}^{\dagger}\right)
\end{equation}
with the coefficients
\begin{eqnarray}
A_{\bf q}  & = & S(2(J_0-J_{\bf q})+A+2J_0'-2J_{\bf q}')+g_L\mu_B H_0 \nonumber
\\ 
&& +S(2A^{xx}_0- A^{zz}_{\bf q} -\sin^2\varphi A^{xx}_{\bf
q}-\cos^2\varphi A^{yy}_{\bf q}\nonumber\\
&&- \sin{2\varphi} A^{xy}_{\bf q}) \\  
B_{\bf q}  & = & S(-A+A^{zz}_{\bf q}-\sin^2\varphi A^{xx}_{\bf
q}-\cos^2\varphi A^{yy}_{\bf q}+\sin{2\varphi} A^{xy}_{\bf q}\nonumber
\\
&&+2i \cos\varphi A_{\bf q}^{yz}-2i\sin\varphi A_{\bf q}^{xz})\, ,
\end{eqnarray}
and the classical ground state energy of the paramagnetic state
\begin{equation}
E_{PM}^{cl} = -NS^2(J_0+J_0' +A^{xx}_0) -g_L\mu_BNSH_0\, , 
\end{equation}
which is independent of the direction of the field. The dispersion relation for
the paramagnetic phase is given by
\begin{equation}
E_{\bf q} = \sqrt{A_{\bf q}^2-|B_{\bf q}|^2}\, .
\end{equation}
In Fig. (\ref{disp-para}) the excitation frequencies are shown for different
field directions ($\varphi=0$). 
When the field is lowered the paramagnetic
phase becomes unstable below a critical field, given by the zero of the
excitation energy ($q_z=0,A_{\bf q}^{yz}=0 $): 
\begin{eqnarray}
  g_L\mu_BH_0^c &=& 2S(J_{\bf q}'-J_0'-A^{xx}_0+\sin^2\varphi
  A^{xx}_{\bf   q}+ \cos^2\varphi A^{yy}_{\bf q}\nonumber \\
&& -\sin{2\varphi}  A^{xy}_{\bf q} )\ .
\label{H_0}
\end{eqnarray}
Considered as a function of ${\bf q}$ the maximum of this expression gives the
critical value (dependent on the angle), and the wave vector ${\bf q}$ reflects
the ordering of the phase below. Assuming a conventional spin-flop phase or an
intermediate phase (two sublattice structure) we would expect the maximum value
for the critical field at wave vectors describing the antiferromagnetic
domains, i.e. ${\bf q}_1$, ${\bf q}_2$ or ${\bf q}_3$. However, the detailed
analysis showed that the paramagnetic phase gets unstable at an incommensurate
wave vector $\tilde {\bf q}(\varphi)$ for certain field directions. Before
considering the general case we study the special angles $\varphi =0^o$ and
$\varphi=90^o$.

\paragraph{$\varphi =0^o$:} 
For this field direction Eq. (\ref{H_0}) simplifies to: 
\begin{equation}
  g_L\mu_BH_0^{cx} = 2S (J_{\bf q}'-J_0'-A^{xx}_0+ A^{yy}_{\bf q})\ .
\end{equation}
This expression is proportional to $V_{\bf q}^{yy}$ defined in
Eq. (\ref{V_0}). The maximum of this component is not achieved for ${\bf q}_1$
but around ${\bf q}_0$. In the limit of vanishing dipole energy the
incommensurate wave vector is given by (compare with Eq. (\ref{inc-wv}))
\begin{equation}
\tilde{\bf q}(0^o)\simeq {\bf q}_0 -1.32\kappa(1,0,0)\, ,
\label{q-tilde}
\end{equation}
i.e. the phase below the paramagnetic phase is an incommensurate phase
resembling a three sublattice structure. Note that this incommensurate wave
vector is different from the wave vector found for the case of vanishing
dipole energy (note the minus sig in Eq. (\ref{inc-wv})). For increasing dipole
energy the wave vector (Eq. (\ref{q-tilde})) changes continuously to  $\bf
q=0$, implying that for $\kappa\to 0$ the system orders ferromagnetically (s.
III).  

For CsNiF$_3$ the critical magnetic field can be evaluated to
\begin{equation}
H_0^{cx} = 290 {\rm mT} \qquad {\rm at} \qquad \tilde {\bf q}(0) = {\pi\over
a}(1.023,0,0)\, .
\end{equation}
The softening of this mode can be seen in Fig. (\ref{disp-para}(a)). This
incommensurate wave vector happens to be near the 
wave vector ${\bf q}_4$, the four-sublattice structure, which describes an 
antiferromagnetic modulation along the  $x$-axis. Note that ${\bf q}_4$ is the
corresponding wave vector at which the N\'eel state becomes unstable (compare
section A).

\paragraph{$\varphi =90^o$:} For magnetic fields parallel to the $y$-axis
the critical value is given by 
\begin{equation}
  g_L\mu_BH_0^{cy} = 2S (J_{\bf q}'-J_0'-A^{xx}_0+ A^{xx}_{\bf q})\, .
\end{equation}
This expression is found to have its maximum value at ${\bf q}_1$,
the antiferromagnetic wave vector of the collinear phase of domain A. For
CsNiF$_3$ this yields
\begin{equation}
H_0^{cy} = 340 {\rm mT} \qquad {\rm at} \qquad \tilde {\bf q}(90^o) = {\bf q}_1
\, .
\end{equation}
At $H_0^{cy}$ the system undergoes a transition into a commensurate phase,
precisely the two sublattice phase corresponding to ${\bf q}_1$. When the
antiferromagnetic exchange energy $J'$ is lowered, the critical field vanishes
for $\kappa \leq 0.20$. This limiting value agrees with the value for the
transition 
between the ferromagnetic and the collinear antiferromagnetic ground state
(Eq. (\ref{kapp-fm})). In contrast to the previous case ($\varphi=0^o$), for
this field direction there does not exist a region for $\kappa$ where an
intermediate phase between ferromagnetic and paramagnetic phase occurs. Owing
to the hexagonal symmetry the critical field for $\varphi =30^o$ is the same as
for $\varphi =90^o$ ($\varphi=-30^o$) but at the antiferromagnetic wave vector
${\bf q}_2$ (${\bf q}_3$) characterizing domain C (B).

\begin{figure}
\centerline{
 \epsfxsize=0.9\columnwidth{\epsfbox{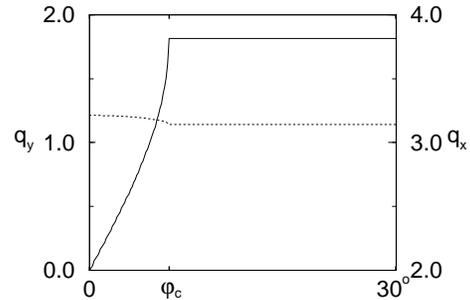}}
}
\vskip .5cm
\caption{Wave vector $\tilde q_y$ and $\tilde q_x$ in the
  paramagnetic phase at which the phase gets unstable depending on the
  direction of the external magnetic field. We used the parameters for
  CsNiF$_3$. $\varphi$ measures the angle between the magnetic field and the
  $x$-axis. For $\varphi > 7.8^o$ the instability appears at point X.}
\label{inkomm-wave}
\end{figure}

\begin{figure}
\centerline{
 \epsfxsize=0.6\columnwidth\rotate[r]{\epsfbox{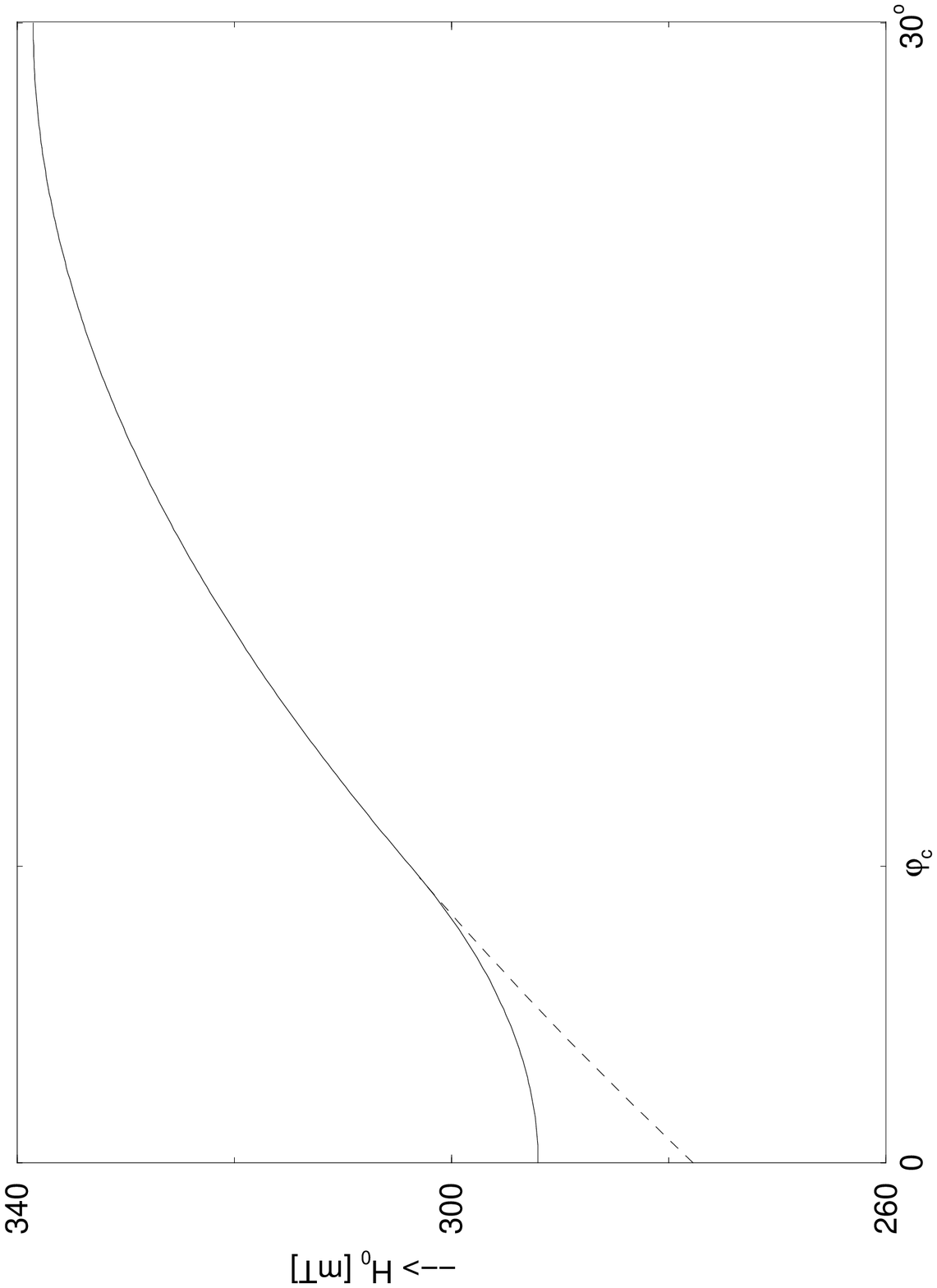}}
}
\vskip .5cm
\caption{Critical field below which the paramagnetic
  phase gets unstable for field direction between $0^o$ and $30^o$. Note that
  there is no discontinuity at $\varphi_c=7.8^o$ but a slight kink. For smaller
  (larger) angles the paramagnetic phase changes to an incommensurate
  (commensurate) phase. The dashed curve indicates the critical field when
  assuming that the soft mode occurs at ${\bf q}_2$. }
\label{inkomm-wave-H_0}
\end{figure}

\begin{figure}
\centerline{
 \epsfxsize=0.8\columnwidth{\epsfbox{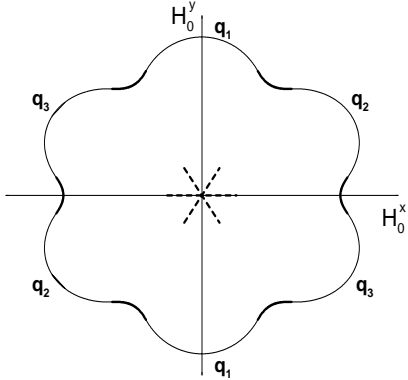}}
}
\vskip .5cm
\caption{Angular dependence of the instability of the
  paramagnetic phase for CsNiF$_3$-like systems. The thick-lined segments on
  the instability curve correspond to the direction of the magnetic field ${\bf
  H}_0 = {\rm H}_0(\cos\varphi,\sin\varphi,0)$, at which the paramagnetic phase
  changes to an incommensurate phase. They are seperated by segments where a
  transition to a commensurate structure happens, especially a two-sublattice
  structure. The corresponding wave vector is given. The dashed lines denote
  the N\'eel state.}
\label{two-inkom}
\end{figure}

\paragraph{$\varphi=$ arbitrary:} 
Finally we turn to arbitrary angles, for which the situation turns out to be
nontrivial. The complete dependence of the field direction $\tilde {\bf q}
(\varphi)$ for CsNiF$_3$ is plotted in Fig. \ref{inkomm-wave}. 
The wave
vector components $\tilde q_x$ and $\tilde q_y$ are plotted as a function of
the angle $\varphi$ for the region in question. A critical angle $\varphi_c
\approx 7.8^o$ is observed above which the wave 
vector is locked to ${\bf q}_2$. Varying the angle $\varphi$ between $0^o$ and
$7.8^o$ the wave vector where the instability sets in changes continuously from
$\tilde {\bf q}(0)$ to $\tilde {\bf q}(\varphi_c)={\bf q}_2$. In
Fig. (\ref{inkomm-wave-H_0}) the 
critical value for the magnetic field is plotted as a function of the angle for
the same angular domain. 
Note that the critical field is a continuous curve
even at the critical angle $\varphi_c$. It has only a small kink at this point.
The dashed curve results from Eq. (\ref{H_0}) when it is assumed that the
instability point occurs at ${\bf q}_2$ for the whole angular
segment. Thus we can see that for $\varphi < \varphi_c$ the incommensurate
structure is favored. Due to the inversion symmetry of the lattice together
with $\tilde {\bf q}(\varphi)$ there is a second modulation wave vector
$-\tilde {\bf q}(\varphi)$. The result of this investigation for a complete
rotation of the magnetic field is summarized in Fig. (\ref{two-inkom}): The
instability line of the paramagnetic phase is shown. There are angular regions
where the paramagnetic structure undergoes a transition to a two-sublattice
structure (annotated by the corresponding wave vector) separated by regions
with incommensurate structures which are plotted with a thick line. The three
dashed lines around zero indicate the N\'eel phases for a longitudinal field. 

\begin{figure}
\centerline{
 \epsfxsize=0.8\columnwidth{\epsfbox{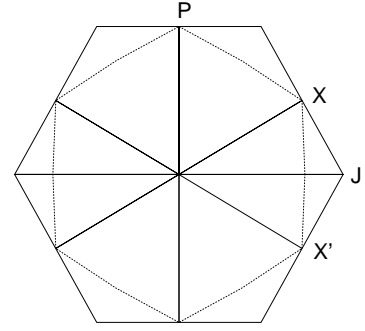}}
}
\vskip .5cm
\caption{Angular dependence of the wave vector $\tilde{\bf
    q}(\varphi)$ on the instability curve from Fig. (10) for
  CsNiF$_3$. Shown is the Brillouin zone from Fig. (6). The wave
  vector happens to be near the edge of the magnetic Brillouin zone for the
  special value of $\kappa=0.79$.}
\label{wave-circ}
\end{figure}

In Fig. (\ref{wave-circ}) the full wave vector dependence within the Brillouin
zone is shown for CsNiF$_3$. Note that the wave vector is locked at the
two-sublattice wave vectors, which cannot be seen in this representation.

\begin{figure}
\centerline{
 \epsfxsize=0.9\columnwidth{\epsfbox{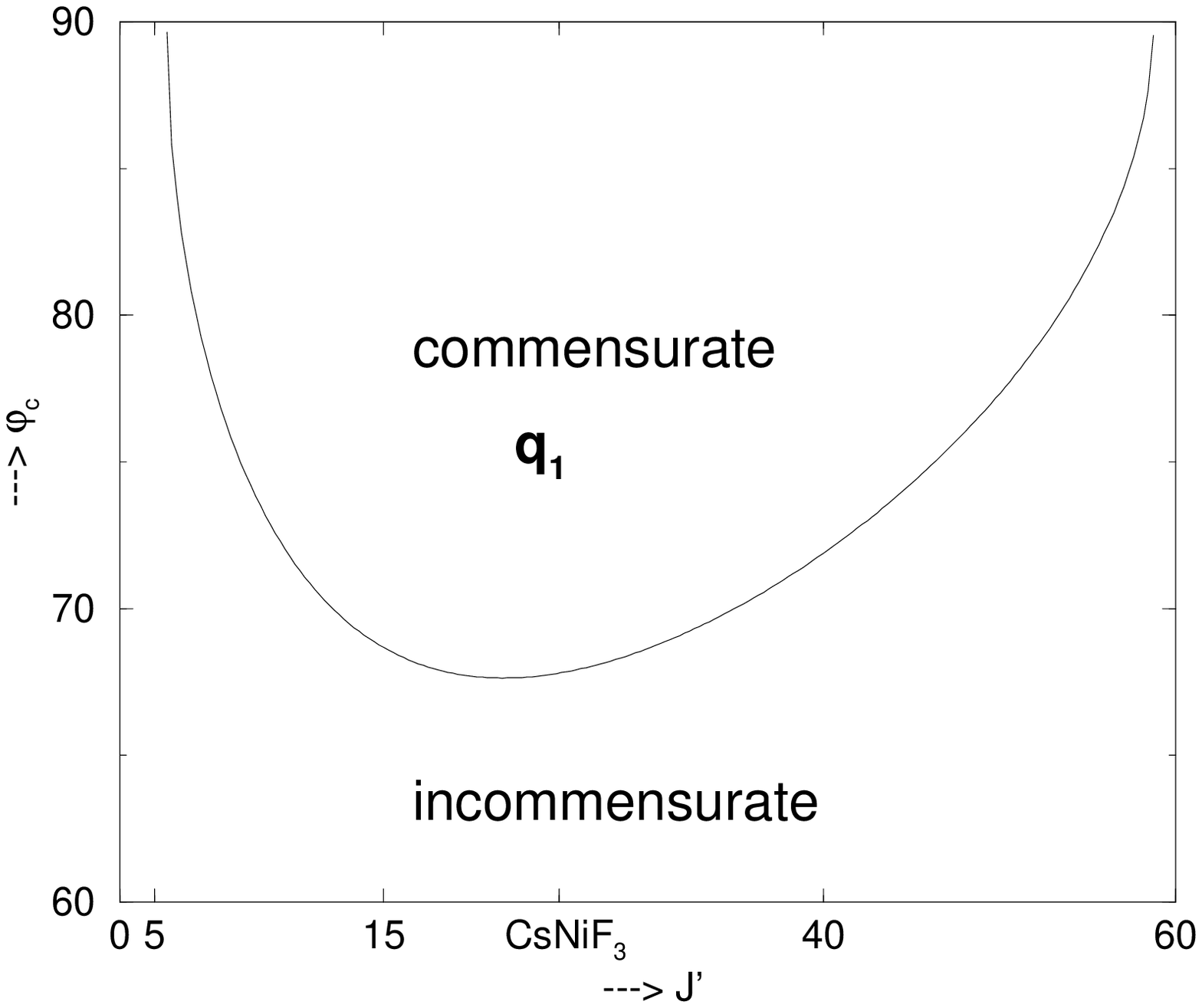}}
}
\vskip .5cm
\caption{The dependence of the critical angle $\varphi_c$ on the
  angtiferromagnetic exchange energy $J'$. The region between the curve and
  $\varphi=90^o$ denotes the commensurate structure right below the instability
  curve.}
\label{phi-c}
\end{figure}

Now we want to derive the critical angle $\varphi_c$, at which the commensurate
and the incommensurate phases coexist, for different strengths of the
antiferromagnetic exchange energy $J'$. This value can be obtained by 
evaluation of the maximum of Eq. (\ref{H_0}) near ${\bf q}_1$. Expanding
the exchange energy and the dipole energy in a Taylor series we get
\begin{eqnarray}
J_{\tilde{\bf q}+{\bf q}_1}' & \approx & J_{{\bf q}_1} +{1\over 2}J'(\tilde
q_x^2-3\tilde q_y^2 )\\
A^{xx}_{\tilde{\bf q}+{\bf q}_1} & \approx & A^{xx}_{{\bf q}_1}-G'(0.816\tilde
q_x^2+0.112\tilde q_y^2) \\
A^{yy}_{\tilde{\bf q}+{\bf q}_1} & \approx & A^{yy}_{{\bf q}_1}+G'(0.816\tilde
q_x^2-0.112\tilde q_y^2) \\
A^{xy}_{\tilde{\bf q}+{\bf q}_1} & \approx & G'1.361\tilde q_x\tilde q_y\, ,
\end{eqnarray}
where we introduced the dipolar strength $G'= G/a_3$ (s. appendix). 
Inserting in Eq. (\ref{H_0}) and setting the determinant of the Hess matrix
zero we find the critical angle as a function of the exchange energy $J'$ which
is shown in Fig. (\ref{phi-c}). 
One can see that the angular segments of the
commensurate structure decrease when the antiferromagnetic exchange energy
takes values near the upper or the lower bound of the collinear
antiferromagnetic phase. For CsNiF$_3$ it happens that the commensurate
segments have nearly the largest possible size.

\subsection{Spin-flop phase, intermediate phase} 

In this section we want to examine the noncollinear phase, i.e. the region
within the instability curve of Fig. (\ref{two-inkom}). In the following
we investigate a magnetic field along the $x$-axis. Concerning domains B and C
it is 
clear that for finite field strength the spins reorientate in order to gain
energy from the interaction with the magnetic field, while domain A remaining
in a N\'eel state cannot gain Zeeman energy. Thus the N\'eel state cannot be a
stable state. In section A we showed via stability calculations that a N\'eel
structure is stabilized up to a critical value (Eq. (\ref{H-crit})). 
Thus the N\'eel state for domain A cannot be a stable state but must be 
metastable. There is an activation energy necessary to flip domain A in a
ground state emerging for domains B or C in the presence of a field, due to the
broken rotation symmetry in the hexagonal plane. The 
situation for a virgin probe is as follows: without magnetic field the system
might be built up of the three domains with equal volume. Raising the magnetic
field (parallel to the spins in domain A) does not change the spin orientation
in domain A but leads to a slight reorientation in domains B and C. Above the
the critical value (Eq. (\ref{H-crit})) the spins in domain A flip (first order
phase transition) to an orientation identical to either domain B or domain
C. For strong magnetic fields one finally enters the paramagnetic phase. When
the magnetic field is decreased thereafter the spins order again in the two
domains, but domain A is never formed again because of the metastability of
this domain with respect to fields along the $x$-direction. The system ends up
in a state where only domains B and C are present. In this scenario we
neglected thermal effects, possible domain wall energies and defects in the
crystal which should be taken into account for a proper description. In fact,
the experiments do not show this scenario; the system always ends in a
three-domain state after a magnetic cycle\cite{STEI74}.

\begin{figure}
\centerline{
 \epsfxsize=0.5\columnwidth\rotate[r]{\epsfbox{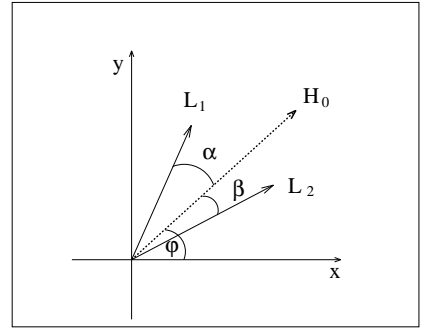}}
}
\vskip .5cm
\caption{Coordinate system for a general two-sublattice spin
  orientation. The spins of sublattice ${\cal L}_1$ (${\cal L}_2$) enclose an
  angle of $\alpha$ ($\beta$) with the magnetic field ${\bf H}_0=H_0
  (\cos\varphi, \sin\varphi , 0)$.}
\label{koordinate}
\end{figure}

For general field direction we expect from the investigation of the instability
line of the paramagnetic phase that in the intermediate phase there are angular
regions with incommensurate modulations separated by region with a commensurate
two-sublattice modulation. Here we want to find out the conditions for
existence of a conventional spin-flop phase, i.e. the ground state consists of
two sublattices on each of which the spin component along the field is the same
($\alpha=\beta$ in Fig. (\ref{koordinate})). 
Due to the large planar anisotropy
$A$ the spins reorientate only within the plane, i.e. we need only one angle
relative to the magnetic field for each spin. A two sublattice structure is
characterized by the wave vectors ${\bf q}_1$, ${\bf q}_2$ and  ${\bf
q}_3$. Introducing (Fig. (\ref{koordinate})) the angle enclosing the $x$-axis
and the magnetic field ${\bf H}_0$ by $\varphi$ and the angle enclosing the
field and the orientation of the spin on the first (second) sublattice ${\cal
L}_1$ (${\cal L}_2$) by $\alpha$ ($\beta$), the ground state energy can be
written in the form:
\begin{eqnarray}
E_{IN}^{cl} &=& -NS^2\{J_0+\cos^2\gamma [J_0'+A_0^{xx}]+\nonumber\\
&&\sin^2\gamma
[J_{{\bf q}}'+\sin^2{(\varphi+\delta)} A_{{\bf q}}^{xx}+\cos^2{(\varphi+
\delta)} A_{{\bf q}}^{yy}\nonumber \\
&& +\sin{2(\varphi+\delta)}A_{{\bf q}}^{xy}]\}-g_L\mu_BNS
H_0\cos\gamma\cos\delta \, ,
\label{Eground}
\end{eqnarray}
where ${\bf q}$ denotes one of the above mentioned two-sublattice wave
vectors. Here we used the more appropriate variables $\gamma$ and $\delta$,
which are related to the original angles by 
\[
\gamma = {{\alpha +\beta}\over 2}, \qquad \delta = {{\alpha -\beta}\over 2}\, .
\]
The minimum value of the ground state energy requires ($\partial_\gamma
E_{IN}^{cl}=0$ and  $\partial_\delta E_{IN}^{cl}=0$):
\begin{eqnarray}
  g_L\mu_BH_0\cos\delta  =  2S[J_{\bf q}'-J_0'-A_0^{xx} +\sin^2{(\varphi
  +\delta)}A_{{\bf q}}^{xx}\nonumber\\
+\cos^2{(\varphi+\delta)} A_{{\bf q}}^{yy}
+\sin{2(\varphi+\delta)}A_{{\bf q}}^{xy}]\cos\gamma\\ 
  g_L\mu_BH_0\cos\gamma\sin\delta  =  S\sin^2\gamma[
  \sin{2(\varphi+\delta)}(A_{{\bf q}}^{xx}-A_{{\bf
  q}}^{yy})\nonumber\\
+2\cos{2(\varphi+\delta)}A_{{\bf q}}^{xy}] \, .
\label{Esf}
\end{eqnarray}
These equations will be discussed now for a spin-flop and an intermediate spin
structure. 

\subsubsection{Spin-flop phase}
For a conventional spin-flop phase ($\alpha=\beta=\gamma$ and $\delta=0$)
these equations can be simplified 
\begin{eqnarray}
g_L\mu_BH_0 & = & 2S[J_{\bf q}'-J_0'-A_0^{xx} +\sin^2\varphi A_{{\bf
q}}^{xx}+\cos^2\varphi A_{{\bf q}}^{yy} \nonumber\\
&&+\sin{2\varphi}A_{{\bf
q}}^{xy}]\cos\gamma \label{sf1}\\
  0 &=& \sin{2\varphi}(A_{{\bf q}}^{xx}-A_{{\bf
  q}}^{yy})+2\cos{2\varphi}A_{{\bf 
      q}}^{xy}\, .
\label{sf2}
\end{eqnarray}
The second relation gives a condition for the existence of a conventional
spin-flop phase. By using the values of the dipole energies and the dipole
relations given in the appendix we find the following solutions for
Eq. (\ref{sf2}):
\begin{center}
\begin{tabular}{|c|ccc|}\hline
  $\varphi$ & ${\bf q}_1$ & ${\bf q}_2$ & ${\bf q}_3$ \\\hline\hline
stable & ${\pi\over 2}, {3\pi\over 2}$ & ${\pi\over 6}, {7\pi\over 6}$ &
${5\pi\over 6}, {11\pi\over 6}$ \\
unstable & $0, \pi$ & ${2\pi\over 3}, {5\pi\over 3}$ &
${\pi\over 3}, {4\pi\over 3}$ \\\hline
\end{tabular}
\end{center}
In the second row the angles are given for the corresponding wave vector. These
solutions represent magnetic field directions perpendicular to the spin
orientation for vanishing field. In the third row we tabulated the solutions
which correspond to field directions along the spin orientation. The latter
solutions represent no minimal ground state energy for exchange energies $J'<
26m$ K. This can be seen from Eq. (\ref{sf1}), where the term in brackets then
is negative. Even for $H_0=0$ and e.g. $\varphi=0^o$ the ground state energy 
\[
E_{IN}^g = -NS^2(J_0+J_{{\bf q}_1}'+A_{{\bf q}_1}^{yy})
\]
is larger than the ground state energy for the N\'eel state
(Eq. (\ref{Eg})). Thus, there is no spin-flop phase for longitudinal fields in
CsNiF$_3$. 

Inserting the values (stable) for the angles from the table in Eq. (\ref{sf1})
we get the static equilibrium relation between the magnetic field and the angle
$\gamma$ 
\begin{equation}
g_L\mu_BH_0  =  2S[J_{\bf q}'-J_0'-A_0^{xx} +A_{{\bf q}_1}^{xx}]\cos\gamma\, .
\end{equation}
Note that this relation holds for all stability values of $\varphi$ due to the
hexagonal symmetry, as can be seen by applying the dipole relations from the
appendix. So we conclude that for most angles $\varphi$ there does not exist a
spin-flop phase except for the above found solutions. 

\subsubsection{Intermediate phase}
In the general case of two independent angles we have to solve equations
(\ref{sf1}) and (\ref{sf2}). As a result one obtains the dependence of the
angles $\alpha(H_0,\varphi)$ and $\beta(H_0,\varphi)$ upon the field strength
and the direction. Since the solution for the general case is nontrivial we
want to study only two special situations, namely $\varphi =0$, $\varphi=90^o$
and a modulation wave vector ${\bf q}_1$ for domain A. In these two cases the
equations of stability (Eqs. (\ref{sf1}) and (\ref{sf2})) can be solved
analytically. For $\varphi=90^o$ we find that the intermediate phase has a
higher ground state energy than the conventional spin-flop phase
($\alpha=\beta$). For $\varphi =0$ the ground state energy of the intermediate
phase exceeds the energy for the N\'eel state for vanishing fields, signifying
the absence of such a state. The complete investigation of the
general case is left for future studies.

\section{Summary}

We have studied the ground state for a planar, quasi one-dimensional
ferromagnet with weak antiferromagnetic exchange and dipole-dipole interaction.
For vanishing exchange energy the spins order ferromagnetically in the
hexagonal plane. With increasing exchange energy the ground state passes
through a collinear phase with three possible domains and switches over to an
incommensurate phase, where the characterizing wave vector moves continuously
from the points $P$, $X$ or $X'$ to the three-sublattice wave vector at point
$J$ (Fig. (\ref{inc-domains})). The three-domain structure of the collinear
phase persists in the incommensurate phase.

Via linear spin wave theory we obtain the dispersion relation for the
ferromagnetic and the collinear antiferromagnetic phase ($H_0=0$). The range
for the collinear antiferromagnetic phase for CsNiF$_3$like systems is
calculated from ground state energy calculations and from stability conditions
concerning to excitations 
\begin{equation}
7 {\rm mK} < J' < 59 {\rm mK}\, ,
\end{equation}
which is consistent with $J'=25$ mK found in recent measurements\cite{baehr}.
The upper and lower bounds are independent of the ferromagnetic exchange energy
$J$ and the planar anisotropy energy $A$, as long as they are much larger than
the dipole energy and the exchange energy $J'$.

In the following we studied the magnetic phase diagram of the collinear
antiferromagnetic phase. For homogeneous external magnetic fields parallel to
the spin orientation the N\'eel phase is stable up to a critical field which
depends only on the dipolar energy (Eq. (\ref{H-crit})). The instability is
shown by the softening of a mode with wave vector ${\bf q}_4$ which describes a
transition to a four-sublattice structure. For magnetic fields transverse to
the spins a reorientation sets in for all finite field values. 

By stability investigations of the paramagnetic phase we obtained a
non-circular instability line for fields in the hexagonal plane. There is a
strong dependence on the field direction on the critical field and the
structure of the phase below the paramagnetic phase. For certain angular
domains the system changes in incommensurate structures which are separated
by commensurate (two-sublattice) structures (Fig. (\ref{two-inkom})). We showed
that the commensurate angular domains are intermediate phases while
conventional spin-flop phases exist only for magnetic field directions
$\varphi=30^o+n60^o$. This is in contrast to the result of Yamazaki et
al.\cite{Yamazaki79}, who found conventional spin-flop phases for all field
directions, even for a longitudinal field where an incommensurate phase should
set in from our calculations. This may result from their semiclassical model
which does not consider the full nature of the dipole-dipole interaction. 

As a result, the magnetic phase diagram reveals an interesting structure of
different phases. The investigation of the complete magnetic phase diagram
(intermediate phase) is left for future studies.

\acknowledgments This work has been supported by the German Federal
Ministry for Research and Technology (BMBF) under the contract number
03-SC4TUM. The work of C.P. has been supported by the Deutsche
Forschungsgemeinschaft (DFG) under the contract no. PI 337/1-1.

\section{Dipole-dipole interaction}
In this section we summarize the most important relations concerning the
dipole-dipole interaction needed in the analysis. We consider a simple
hexagonal lattice, with lattice constant $a$ in the triangular plane and
lattice constant $c$ perpendicular to the plane. The direct lattice is
paramatrized by 
\[
{\bf x}_l = a(l_1,0,0)+{a\over 2}(l_1,\sqrt 3l_2,0)+c(0,0,l_3),
\]
\[\qquad l_i = 
0,\pm1,\pm2,...
\]
and the reciprocal lattice by
\[
{\bf G} = {2\pi\over a}(h_1,0,0)+{2\pi\over \sqrt 3a}(2h_2-h_1,0)+ {2\pi\over
c}(0,0,h_3),
\]
\[ h_i = 0,\pm1,\pm2,...
\]
The Fourier transform of the
dipole energy (Eq. (\ref{Dipoltensor})) 
\[
A^{\alpha\beta}_{\bf q} =
\]
\begin{equation}
 -G\lim_{{\bf x}\to 0}{1\over 2}\left(\sum_le^{i{\bf q
x}_l}\left({\delta^{\alpha\beta}\over 
|{\bf x -x}_l|^3}-{3x_l^{\alpha}x_l^{\beta} \over |{\bf
x-x}_l|^5}\right)-{1\over |{\bf x}|^3}\right)
\end{equation}
with 
\[
G=(g_L\mu_B)^2
\]
is found by the Ewald summation technique\cite{Bonsal77}. There the summation
is split into a part over the direct lattice and a part over the reciprocal
lattice and, by using in an intermediate step the generalized
theta function\cite{Bonsal77}, we derive the following relation ($a_3 = {\sqrt
3\over 2}a^2c$): 
\begin{eqnarray*}
A^{\alpha\beta}_{\bf q} \,{a_3\over G} &=& {2\pi\over 3}\delta^{\alpha\beta}
-\pi\delta^{\alpha\beta}\sum_l'e^{i{\bf q x}_l}\varphi_{1/2}
\left({\pi\over a_3^{2/3}} |{\bf x}_l|^2\right)\\
&&+{2\pi^2\over a_3^{2/3}}\sum_l' e^{i{\bf q x}_l} x_l^\alpha
x_l^\beta\varphi_{3/2}\left({\pi\over a_3^{2/3}} |{\bf x}_l|^2 
\right)
\end{eqnarray*}
$$-{a_3^{2/3}\over 2}\sum_{\bf G}({\bf q}+{\bf G})^\alpha
({\bf q}+{\bf G})^\beta\varphi_0\left({a_3^{2/3}\over 4\pi}|{\bf q}+{\bf G}|^2
\right)\, .
$$
Here $l$ refers to the direct lattice and ${\bf G}$ to the
reciprocal lattice. The prime excludes the value $l= 0$ and we used the Misra
functions
$$
\varphi_n(x)=\int_1^\infty dt\, t^ne^{-xt}\, .
$$
Expanding the dipole tensor for small wave vectors yields:
\begin{eqnarray}
A^{\alpha\beta}_{\bf q} {a_3\over G} &=& 
-2\pi\left({q^\alpha q^\beta\over q^2}-{\delta^{\alpha\beta}\over 3}\right)-
\pi\delta^{\alpha\beta}\sum_l'\varphi_{1/2}
\left({\pi\over a_3^{2/3}} |{\bf x}_l|^2\right)\nonumber\\
&&+{2\pi^2\over a_3^{2/3}}\sum_l' 
x_l^\alpha x_l^\beta\varphi_{3/2}\left({\pi\over a_3^{2/3}} |{\bf x}_l|^2
\right)\nonumber\\
&&+\sum_{\bf G}'G^\alpha G^\beta \varphi_0\left(a_3^{2/3}|{\bf G}|^2\over
4\pi\right)+{\cal O}(q^\alpha q^\beta)\\
&=& -2\pi\left({q^\alpha q^\beta\over q^2}-{\delta^{\alpha\beta}\over 3}\right)
+a_{lat}\, ,
\label{A_0}
\end{eqnarray}
where we have introduced a lattice dependent term $a_{lat}$. This expression is
non-analytic due to the semiconvergence of the dipole sum in three
dimensions. The value at ${\bf q}=0$ depends on the direction of the limiting
process expressing the shape dependence of the system. 
For a cubic lattice the lattice dependent term vanishes $a_{lat}=0$, so the
dipole energy for a ferromagnetic structure is zero. In contrast, this term
does not vanish for a hexagonal lattice. For non-spherical surfaces this
equation has to be corrected by the demagnetization factor. 

In the following table the dipole components are listed for selected wave
vectors in units of $G/a_3$ and for $k=a/c=2.39$, the ratio of the lattice
constants for CsNiF ($q_z=0$)
\bigskip
\begin{center}
\begin{tabular}{|c|cccccc|}
\hline
& ${\bf q}=0$ & ${\bf q}_0$ & ${\bf q}_1$ & ${\bf q}_2$ & ${\bf q}_3$ & ${\bf
q}_4$  \\
\hline
$A^{xx}_{\bf q}$ & -2.805 & -5.946 & -3.392 & -7.223 & -7.223 &-7.520 \\
$A^{yy}_{\bf q}$ & -2.805 & -5.946 & -8.501 & -4.669 & -4.669 & -4.373\\
$A^{zz}_{\bf q}$ &  9.798 & 11.893 & 11.893 & 11.893 & 11.893 & 11.893\\
$A^{xy}_{\bf q}$ & 0 & 0 & 0 & -2.212 & 2.212 &0\\
\hline
\end{tabular}
\end{center}
\vskip .5cm
Due to the sixfold rotation symmetry of the lattice, we can derive a general
relation between dipole components in different directions:
\begin{eqnarray}
4A_{{\bf q}(\varphi)}^{yy} & = & 3 A_{{\bf q}(\varphi +60^o)}^{xx}
+ A_{{\bf q}(\varphi +60^o)}^{yy} -2\sqrt 3A_{{\bf q}(\varphi
+60^o)}^{xy}\, .
\end{eqnarray}
${\bf q}(\varphi)$ denotes a wave vector along direction $\varphi$ in the
Brillouin zone. From this formula we obtain relations for different points in
the Brillouin zone:
\begin{eqnarray}
4A_{{\bf q}_3}^{yy} & = & 3 A_{{\bf q}_2}^{xx}
+A_{{\bf q}_2}^{yy} -2\sqrt 3 A_{{\bf q}_2}^{xy}\\
4 A_{{\bf q}_1}^{xx} & = & A_{{\bf q}_2}^{xx}+3A_{{\bf q}_2}^{yy}-2\sqrt
3A_{{\bf q}_2}^{xy}\\
4A_{{\bf q}_1}^{xx} & = & A_{{\bf q}_3}^{xx}+3A_{{\bf q}_3}^{yy}+2\sqrt
3A_{{\bf q}_3}^{xy} \\
0& = & 3(A_{{\bf q}_2}^{xx}-A_{{\bf q}_2}^{yy})+2\sqrt 3 A_{{\bf q}_2}^{xy}\,
. 
\label{dipol-rel}
\end{eqnarray}

\end{document}